\documentclass[twocolumn,preprintnumbers,amsmath,amssymb,prb]{revtex4}

\usepackage{graphicx}% Include figure files
\usepackage{amssymb,amsmath}
\usepackage{dcolumn}% Align table columns on decimal point
\usepackage{bm}% bold math
\usepackage{color}
\usepackage{enumerate}
\usepackage{epstopdf}

\begin{document}

\title{Enhancement of second-harmonic generation in micropillar resonator due to the engineered destructive interference}
\author{S. A. Kolodny$^{1}$}\email{s.kolodny@metalab.ifmo.ru}
\author{V.K. Kozin$^{1,2}$}
\author{I. V. Iorsh$^{1}$}

\affiliation{$^1$ITMO University, Kronverkskiy prospekt 49, Saint Petersburg 197101, Russia} \affiliation{$^2$Science Institute, University of Iceland, Dunhagi-3, IS-107 Reykjavik, Iceland}

\begin{abstract}
We show, that engineering the cavity aspect ratio of AlGaAs micropillar  resonator allows for the  two order of magnitude increase in the efficiency of second harmonic generation at telecom wavelengths.
\end{abstract}

\maketitle

Resonant  photonic nanostructures based on high refractive index materials have proven to be a highly efficient tool to enhance the efficiency of nonlinear optical processes~\cite{sain2019nonlinear}, such as e.g. second~\cite{carletti2016shaping}, third~\cite{shcherbakov2014enhanced} and higher harmonic generation~\cite{liu2018all}. The advantages of these systems over the bulk set-ups and plasmonic nanostructures are the lift of the phase matching conditions~\cite{kauranen2013freeing} and much lower material losses~\cite{baranov2017all}, respectively. At the same time, due to the subwavelength mode volume (and thus relatively short time of the interaction of light with the nonlinear medium as compared to the bulk structures) and the weaker field localization as compared to the plasmonic structures, the all-dielectric set-ups face a challenge of enhancing the efficiency of non-linear processes. The key to enhance the efficiency is to incorporate nonlinear materials onto a nano-scale high quality factor $Q$ resonator, where light can be stored in a small volume $V$, since the efficiency of the nonlinear conversion is proportional to the ratio $Q/V$~\cite{vahala2003optical}. The seminal example of such kind of structures are photonic crystal cavities which demonstrate large values of nonlinear frequency conversion efficiency~\cite{buckley2014second}. At the same, photonic crystal cavities demand sophisticated fabrication techniques, since their quality factor is extremely sensitive to the geometry imperfections.

In contrast, isolated subwavelength semiconductor nanoantennae are fairly easy to fabricate. At the same time, their quality factor defined by the low order electric and magnetic multipole resonances is typically relatively low. However, as has been recently shown, engineering of the shape of the nanoantenna allows to achieve the destructive interference of the low order multipole modes~\cite{KoshelevHighQ,koshelev2019nonradiating}. As a result these structures may support high quality optical modes, characterized at the same time by relatively small mode volumes. Since, the effect of the emergence of the dark modes due to the destructive interference is analogous to the bound states in the continuum (BIC) arising in periodic structures~\cite{hsu2016bound}, these modes are usually referred to as quasi-BIC states. It has been shown experimentally that these states, supported by the AlGaAs pillars facilitate substantial increase in the second harmonic generation efficiency~\cite{KoshelevScience2020}.
\begin{figure}[!h]
\includegraphics[width=\linewidth]{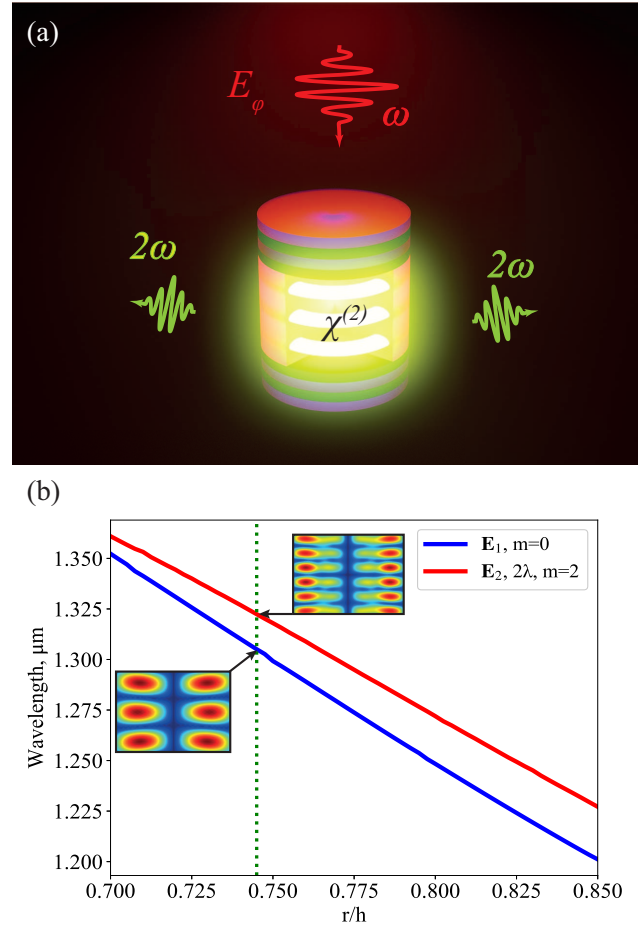}
\caption{(a) The sketch of the system under consideration. (b) Dependence of the wavelength of the modes on the $r/h$ ratio of the cavity. The wavelength of the mode $\mathbf{E}_2$  is doubled for better comparison (to demonstrate that $\omega_2$ is close to $2\omega_1$).}\label{E1E2}
\end{figure}

At the same time, even at the quasi-BIC regime, individual semiconductor pillars are characterized by fairly modest quality factors in mid infrared and optical frequency ranges. The quality factor can be substantially increased if the pillar is sandwiched between Bragg reflectors, which suppress the radiation losses through the top and bottom of the pillar. The resulting structures, pillar microcavities, is conventionally characterized by large values of $Q/V$ ratio and are routinely used to enhance the light-matter interactions at the nanoscale~\cite{lecamp2007submicron,lin2016cavity}.

The main source  of the radiation losses in pillar microcavities is due to the radiation leakage through the sidewalls, which increase as the diameter of the cavity is decreased. We have recently shown, that the at certain ratios of cavity radius to cavity height, the destructive interference occurs similar to the one in the quasi-BIC state, which suppresses the side-wall leakage and resonantly increases the quality factor while preserving the effective mode volume~\cite{kolodny2020q}. Here, we show that this quasi-BIC state occurring in pillar microcavities can be used to substantially increase the efficiency of the second harmonic generation.

We investigate the second-harmonic generation (SHG) in a micropillar resonator with radius 500 nm, consisting of an AlGaAs cylinder, sandwiched between two DBR mirrors (GaAs-AlGaAs, 30 layers on top and bottom), as shown in Fig.~\ref{E1E2}(a). The micropillar resonator is tuned to the quasi-BIC regime~\cite{KolodnyIorshBIC, KoshelevHighQ}. Refractive indices used in study are 3.48 and 3.08 for GaAs\cite{skauli2003improved} and AlGaAs\cite{adachi1989optical}, respectively, while their imaginary parts are neglected since they are at least two orders of magnitude smaller.

The cavity is placed in the background field (pump), which in our case is supposed to be a superposition of two linearly polarized Hermite-Gauss beams \cite{zhan2009cylindrical} that result in an azimuthally polarized field with the azimuth number $m = 0$: 
\begin{equation}
    \mathbf{E}_{\mathrm{b}}(\omega,\mathbf{r}) = \mathrm{HG}_{01}( \omega,\mathbf{r}) \mathbf{e}_x + \mathrm{HG}_{10}( \omega,\mathbf{r} ) \mathbf{e}_y.
\end{equation}
The AlGaAs has a non-vanishing tensor of the second-order nonlinear susceptibility $\chi_{i j k}^{(2)}$. This tensor contains only off-diagonal elements in the
principal axis system of the zinc blende crystalline structure~\cite{Boyd}, with the components being non-zero only if $i\neq j\neq k$, $\chi_{xyz}^{(2)} \equiv \chi^{(2)}=290 \mathrm{pm} / \mathrm{V}$.

In our analysis we focus on the two modes of the pillar: $\mathbf{E}_{1,2}$ with the real and imaginary parts of the eigenfrequencies being equal to $\omega_{1,2}$ and $\gamma_{1,2}$, respectively. The pillar is pumped at the frequency $\omega$ close to $\omega_1$, and the frequency $\omega_2$ of the mode $\mathbf{E}_2$ is assumed to be close to $2\omega$ as shown in Fig.~\ref{E1E2}(b). So in our simulations we consider TE$_{012}$ mode as $\mathbf{E}_1$ since it has a confirmed quasi BIC, and the TE$_{215}$ mode as $\mathbf{E}_2$ because of the proximity of $\omega_2$ to $2\omega$ (the modes field spatial distributions are shown as insets on the Fig.~\ref{E1E2}(b)). The cavity radius is fixed in our study and we vary only its height as well as the Bragg layers period to make the center of the bandgap tuned to the mode frequency.
\begin{figure}[!h]
\centering\includegraphics[width=\linewidth]{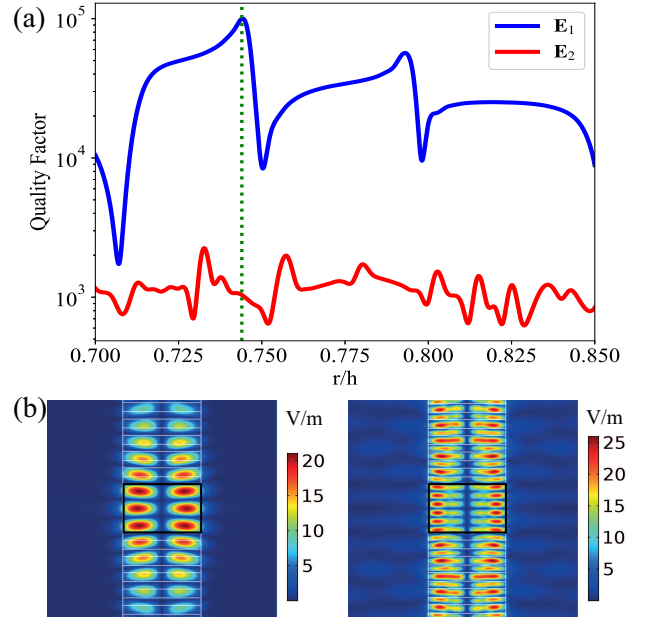}
\caption{(a) The quality factor of the modes $\mathbf{E}_1$ and $\mathbf{E}_2$. The vertical line represents the position of the highest value of $Q$ at r/h = 0.745. (b) The modes field distributions, left picture corresponds to $\mathbf{E}_1$, right picture corresponds to $\mathbf{E}_2$. Note that only part of DBR is shown on pictures while whole number of DBR periods is 30 on top and bottom.}\label{Q1Q2}
\end{figure}
Our main goal is to calculate the nonlinear conversion coefficient, showing the efficiency of the second-harmonic generation, and defined as the ratio between the total SHG power and the pump power squared: $P(2\omega) / P_0(\omega)^2$. The corresponding expression for the total SHG power is given by~\cite{KoshelevScience2020}
\begin{equation}
    P(2 \omega)=\frac{8\pi}{c}\left(\frac{2\omega}{c}\right)^2 \kappa_{2} Q_{2} L_{2}(2 \omega) \kappa_{12}\left[Q_{1} L_{1}(\omega) \kappa_{1}(\omega) P_0(\omega)\right]^{2}\label{totalSH}.
\end{equation}
Here $Q_j=\omega_j/(2\gamma_j)$ is the mode quality factor. The spectral overlap factor is defined as
\begin{equation}
L_{j}(\omega)=\frac{\gamma_{j}^{2}}{\left(\omega-\omega_{j}\right)^{2}+\gamma_{j}^{2}},
\end{equation}
and the so-called coupling $\kappa_1$
\begin{equation}
    \kappa_{1}(\omega)=\frac{\left|(\omega / c) \int d^3 \mathbf{r}^{\prime} \Delta \epsilon\left(\omega, \mathbf{r}^{\prime}\right) \mathbf{E}_{1}\left(\mathbf{r}^{\prime}\right) \cdot \mathbf{E}_{\mathrm{b}}\left(\omega, \mathbf{r}^{\prime}\right)\right|^{2}}{\left(2 \gamma_{1} / c\right) N_{1} (8\pi/c)P_0(\omega)},
\end{equation}
cross-coupling $\kappa_{1,2}$
\begin{equation}
    \kappa_{12}=\frac{\left|\sum_{i, j, k} \chi_{i j k}^{(2)} \int d^3 \mathbf{r} E_{2, i}(\mathbf{r}) E_{1, j}(\mathbf{r}) E_{1, k}(\mathbf{r})\right|^{2}}{\left(N_{2} \omega_{2} / c\right)\left(N_{1} \omega_{1} / c\right)^{2}}
\end{equation}
and decoupling $\kappa_2$ coefficients
\begin{equation}
    \kappa_{2}=\frac{\oint_{S_{\text{pillar}}} d \mathbf{S} \cdot \operatorname{Re}\left[\mathbf{E}_{2} \times \mathbf{H}_{2}^{*}\right]}{\left(2 \gamma_{2} / c\right) N_{2}},
\end{equation}
where they are expressed in terms of the spatial mode profiles $\mathbf{E}_{1,2}(\mathbf{r})$ and the pillar parameters. $N_{1,2}$ are the normalisation constants for the modes $\mathbf{E}_{1,2}$ (see Ref.~\cite{PhysRevA.90.013834}), respectively, and $\Delta \epsilon(\omega, \mathbf{r})=\epsilon(\omega, \mathbf{r})-\epsilon_{\mathrm{b}}(\omega, \mathbf{r})$ is the difference between the resonator dielectric function and the background dielectric function. It should be noted that in our calculations the second harmonic generation is considered only in the AlGaAs. 

\begin{figure}[!h]
\includegraphics[width=\linewidth]{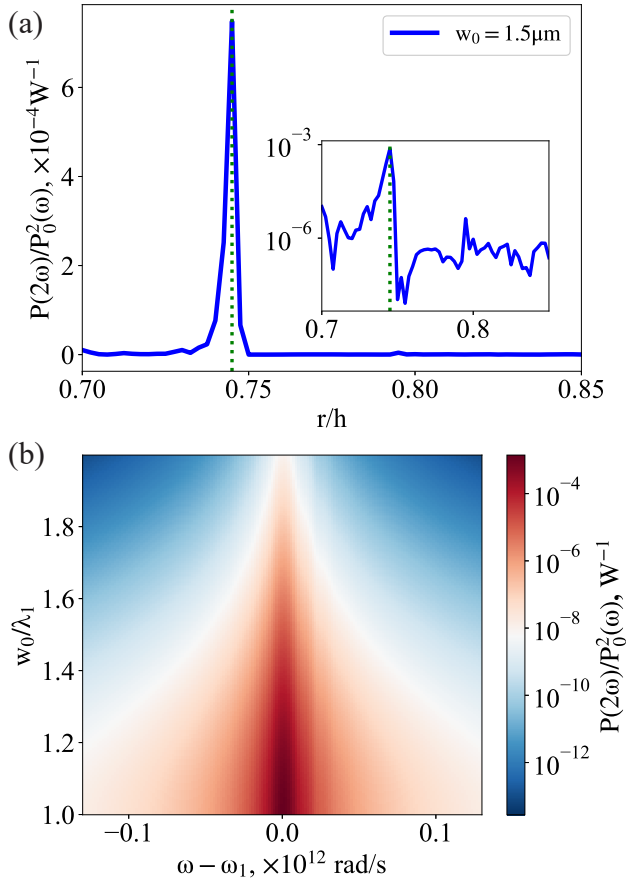}
\caption{(a) Dependence of the nonlinear conversion coefficient on the aspect ration $r/h$ of the cavity for the beam waist radius $w_0=1.5$~$\mu$m. Inset represents the same plot but in the logarithmic scale. (b) Dependence of the nonlinear conversion coefficient on the beam waist radius $w_0$ and the frequency of the background field $\omega$. The highest value is about $1.4\times 10^{-3}$~W$^{-1}$.}\label{n_rh}
\end{figure}

The $Q$-factor of the $\mathbf{E}_1$ mode reaches its highest value $Q \approx 10^{5}$ when the aspect ratio $r/h$ of AlGaAs cavity is approximately equal to 0.745, see Fig.~\ref{Q1Q2}(a). The large value of the quality factor justifies the neglect of non-radiative losses. As one can see the quality factor has several maxima in the considered region of aspect ratios. We note, that the oscillatory behaviour of the quality factor of the pillar microcavity as a function of aspect ratio has been observed experimentally~\cite{lalanne2004electromagnetic}. This effect arises because of the low contrast of Bragg reflectors when the $\mathbf{E}_1$ mode interacts with low-quality modes of the Bragg reflectors. In case of high contrast Bragg reflectors such low-quality modes do not arise and the quality factor of $\mathbf{E}_1$ has a single maximum~\cite{kolodny2020q} at the aspect ratio $r/h=0.745$. In the same region of the aspect ratio the mode $\mathbf{E}_2$ has a very oscillating quality factor since the center of Bragg reflectors bandgap is tuned to the $\mathbf{E}_1$ frequency and it can be seen from field distributions shown in Fig.~\ref{Q1Q2}(b). While the $\mathbf{E}_1$ field quickly fades in the Bragg AlGaAs/GaAs layers, the $\mathbf{E}_2$ field is almost not influenced by the reflectors. Since the $Q$-factor of the $\mathbf{E}_1$ mode is included in the expression for total SHG power (Eq.~\ref{totalSH}) in the second power, it is better to consider the point of the highest $Q$-factor of the $\mathbf{E}_1$ mode, where due to this fact the SHG power will achieve the highest values. 

The dependence of second harmonic nonlinear coefficient on the aspect ratio $r/h$ was studied with a fixed beam waist radius equal to 1.5 $\mu$m (see Fig.~\ref{n_rh} (a)). As it can be seen from this plot, the nonlinear coefficient has a pronounced maximum at $r/h=0.745$, where the coefficient is at least an order of magnitude larger in comparison with the rest of the aspect ratio area (see the inset on Fig.~\ref{n_rh}(a)), and it is about $7.53\times 10^{-4}$~W$^{-1}$. At this point the nonlinear coefficient dependence on the background field frequency in the frequency domain looks like a narrow peak ($\sim 1\times 10^{10}$ rad/s) because of the small value of $\gamma_1$, which enters the spectral overlap factor $L_1(\omega)$.  

As a next step, the dependence of the nonlinear coefficient on the beam waist radius of the background field was studied (see Fig.~\ref{n_rh}(b)). The maximal achieved nonlinear coefficient $\approx 1.4\times 10^{-3}$~W$^{-1}$ at the value of $w_0$ equal to the wavelength of the $\mathbf{E}_1$  mode, corresponds to $r/h=0.745$. Considering shorter wavelengths for such geometry requires taking into account additional effects such as generation of higher harmonics and difference frequency generation.

To conclude, in this work we have investigated the second harmonic generation in a micropillar AlGaAs/GaAs resonator, and have shown that it gets significant enhancement in the quasi-BIC regime. Compared to a single resonator\cite{KoshelevScience2020}, the achieved theoretical values are higher by at least an order of magnitude, despite the fact that the $Q$-factor is much higher ($10^5$ versus $10^2$). Here, the decisive factor is the mode volume, which is larger for the micropillar resonator (especially for the $\mathbf{E}_2$ mode), which leads to a decrease in the effective generation of the second harmonic. The crucial role here is played by the refractive index contrast of the Bragg reflectors: Bragg reflectors with a high contrast allow to achieve higher values of the quality factor of the $\mathbf{E}_1$ mode and avoid additional modes in the Bragg layers. On the other hand, since the modes hybridization occurs outside the resonant cavity such additional modes may be used to create an artificial BIC for the modes of interest, improving its nonlinear properties. Thus we believe that the the presented results can be applied in problems of nonlinear nanophotonics and in the practical implementation of quantum devices, where high nonlinearity plays an important role.

The authors would like to thank Kirill Koshelev from ITMO University and Sergei Yankin from COMSOL.This work was supported by the Ministry of Science and Higher Education of Russian Federation, goszadanie no. 2019-1246.

% Bibliography


\begin{thebibliography}{10}

\bibitem{sain2019nonlinear}
B. Sain, C. Meier, and T. Zentgraf.
\newblock {\em Advanced Photonics}, 1(2):024002, 2019.

\bibitem{carletti2016shaping}
L. Carletti, A. Locatelli, D. Neshev, and C. De~Angelis.
\newblock {\em ACS Photonics}, 3(8):1500--1507, 2016.

\bibitem{shcherbakov2014enhanced}
M.~R. Shcherbakov, D.~N. Neshev, B. Hopkins, et~al.
\newblock {\em Nano letters}, 14(11):6488--6492, 2014.

\bibitem{liu2018all}
S. Liu, P.~P. Vabishchevich, A. Vaskin, et~al.
\newblock {\em Nature communications}, 9(1):1--6, 2018.

\bibitem{kauranen2013freeing}
M. Kauranen.
\newblock {\em science}, 342(6163):1182--1183, 2013.

\bibitem{baranov2017all}
D.~G. Baranov, D.~A. Zuev, S.~I. Lepeshov, et~al.
\newblock {\em Optica}, 4(7):814--825, 2017.

\bibitem{vahala2003optical}
K.~J. Vahala.
\newblock {\em nature}, 424(6950):839--846, 2003.

\bibitem{buckley2014second}
S. Buckley, M. Radulaski, J. Petykiewicz, et~al.
\newblock {\em ACS Photonics}, 1(6):516--523, 2014.

\bibitem{KoshelevHighQ}
M.~V. Rybin, K.~L. Koshelev, Z.~F. Sadrieva, et~al.
\newblock {\em Phys. Rev. Lett.}, 119:243901, Dec 2017.

\bibitem{koshelev2019nonradiating}
K. Koshelev, G. Favraud, A. Bogdanov, et~al.
\newblock {\em Nanophotonics}, 8(5):725--745, 2019.

\bibitem{hsu2016bound}
C.~Wei Hsu, B.~Zhen, A.~Douglas Stone, et~al.
\newblock {\em Nature Reviews Materials}, 1(9):1--13, 2016.

\bibitem{KoshelevScience2020}
K. Koshelev, S. Kruk, E. Melik-Gaykazyan, et~al.
\newblock {\em Science}, 367(6475):288--292, 2020.

\bibitem{lecamp2007submicron}
G. Lecamp, J.-P. Hugonin, P. Lalanne, et~al.
\newblock {\em Applied Physics Letters}, 90(9):091120, 2007.

\bibitem{lin2016cavity}
Z. Lin, X. Liang, M. Lon{\v{c}}ar, et~al.
\newblock {\em Optica}, 3(3):233--238, 2016.

\bibitem{kolodny2020q}
S. Kolodny and I. Iorsh.
\newblock In {\em Journal of Physics: Conference Series}, volume 1461, page
  012067. IOP Publishing, 2020.

\bibitem{KolodnyIorshBIC}
S. Kolodny and I. Iorsh.
\newblock {\em Opt. Lett.}, 45(1):181--183, Jan 2020.

\bibitem{skauli2003improved}
T.~Skauli, P.S.~Kuo, K.L.~Vodopyanov, et~al.
\newblock {\em Journal of Applied Physics}, 94(10):6447--6455, 2003.

\bibitem{adachi1989optical}
S. Adachi.
\newblock {\em Journal of Applied Physics}, 66(12):6030--6040, 1989.

\bibitem{zhan2009cylindrical}
Q. Zhan.
\newblock {\em Advances in Optics and Photonics}, 1(1):1--57, 2009.

\bibitem{Boyd}
R.~W. Boyd.
\newblock Academic, 2003.

\bibitem{PhysRevA.90.013834}
M.~B. Doost, W.~Langbein, and E.~A. Muljarov.
\newblock {\em Phys. Rev. A}, 90:013834, Jul 2014.

\bibitem{lalanne2004electromagnetic}
P.~Lalanne, J.-P. Hugonin, and J.-M. G{\'e}rard.
\newblock {\em Applied Physics Letters}, 84(23):4726--4728, 2004.

\end{thebibliography}
\end{document}